\documentclass{ws-procs9x6}

\begin{document}
\begin{flushleft}
KCL-PH-TH/2011-35 \\
LCTS/2011-19 \\
CERN-PH-TH/2011-276
\end{flushleft}

\title{RECENT RESULTS FROM INDIRECT  AND DIRECT DARK MATTER SEARCHES: 
THEORETICAL SCENARIOS}

%\author{A. B. AUTHOR$^*$ and C. D. AUTHOR}
%
%\address{University Department, University Name,\\
%City, State ZIP/Zone, Country\\
%$^*$E-mail: ab\_author@university.com\\
%www.university\_name.edu}

\author{N. E. MAVROMATOS}

\address{Theoretical Particle Physics and Cosmology Group, Department of Physics, \\
King's College London, Strand, London, WC2R 2LS, UK; \\
CERN, Physics Department, Theory Division, CH-1211 Geneva 23, Switzerland. }

\begin{abstract}
In this review, I  discuss briefly theoretical scenarios concerning the
interpretation of recent results from indirect and direct dark matter searches, with emphasis on the former.

\end{abstract}


\bodymatter

\section{Introduction} 

There is current evidence from a plethora of astrophysical measurements that the energy budget of our Universe consists of more than 70\% of a mysterious
Dark Energy component, responsible for its current accelerating expansion, and another 23\% of Dark Matter (DM), also of unknown origin.

In this article I will concentrate on theoretical interpretations of recent results from indirect DM searches,  that is excess of $\gamma$-rays and neutrinos from galactic sources or the Sun above the expected cosmic backgrounds,  as well as matter-antimatter asymmetries (positron excess) around the Earth observed recently in the cosmic ray (CR) spectrum by PAMELA and confirmed by FERMI~\cite{fermi}. 

The structure of this article is as follows: 
In the next section \ref{sec:susy}, I review the properties of DM candidates in 
supersymmetric(SUSY)/supergravity(SUGRA) models, placing  the emphasis on the  (significant) dependence of the various predictions (in particular with relevance to indirect DM searches) on the specific theoretical model used. This discussion has particular relevance these days,  where LHC results seem to disfavour large parts of the parameter space of simplest SUSY models. In sec. \ref{sec:pam} I explain how WIMP models (including SUSY ones) can accommodate the results on positron excess in Cosmic Ray spectra. In sec. \ref{sec:sterile}, I discuss sterile neutrinos as DM candidates in non supersymmetric models. In sec. \ref{sec:simp}, I describe other interesting particle physics candidates of DM, charged or neutral (fermionic), which may be strongly interacting with SM particles (Strongy Interacting Massive Particles (SIMP)). Finally, in section \ref{sec:otherDM}, I discuss other ideas on DM, including axions as well as the possibility of having Dark Atoms in non supersymmetric extensions of the SM, or mediation of the interactions of DM with SM particles via the exchange of $Z'$ gauge bosons, pertaining to extra $U(1)' $ gauge groups.

\section{ Supersymmetry/Supergravity and indirect DM searches \label{sec:susy}}

The most extensively studied model so far, from the point of view of supersymmetry searches at colliders and in particular LHC, 
is the five-parameter Constrained Minimal Supersymmetric Standard Model  (CMSSM) (and its minimal Supergravity (mSUGRA) variant)  with R-symmetry conservation.  The Cold DM  candidate in this class of models is the neutralino, which is the Lightest SUSY Particle (LSP) in the spectrum and hence stable. 

Indirect searches for neutralinos $\chi$ are motivated by the fact that neutralino annihiliation in the galaxies produces gamma ray excess in the relevant spectra, and this constitute a signal for this type of DM. 
Although there are attempts to provide model independent fits to such photon spectra~\cite{fits},  nevertheless due to the weakness of the signal there is significant sensitivity to the particular theoretical model for DM, as we now discuss. The most studied example are the photon spectra from neutralino $\chi$ annihilation at the core of our Galaxy. The total annihilation cross section rates of the mSUGRA or CMSSM have been studied in ref.~\refcite{spanosgamma} and the photon spectra from the core of our galaxy due to LSP ($\chi$) annihilation 
have been estimated in the region of the parameter space of the models that are compatible with the WMAP constraints: (i) the stau $\overline{\tau}_1 -\chi$ co-annihilation strip, (ii) the focus point region (in which $\chi$ has an enhanced Higgsino component) and (iii) the funnel at large tan$\beta$, in which the annihilation rate is enhanced  by poles of nearby heavy MSSM Higgs bosons. The important point to notice is that, as the relevant calculations show, annihilation attenuates rapidly  with decreasing tan$\beta$ and 
increasing m$_{1/2}$. The analysis of ref.~\refcite{spanosgamma} has been quite thorough, involving detailed calculations of WMAP compatible branching fractions of $\chi-\chi$ annihilation into SM particle pairs at certain characteristic CMSSM benchmark points. The resulting CMSSM total 
$\gamma$-ray flux  as a function of the energy threshold  has been computed.  The prospects for detection depend crucially on the astrophysical $\gamma$-ray background, which has three known components so far: (a) diffuse galactic emission (DGE), from nucleon-nucleon interactions producing $\pi^0 $ which subsequently decay to gamma rays, and electron bremsstrahlung as it it scattered by a nucleus, (b) Isotropic Extragalactic (possibly) Contributions (IGRB) from a plethora of sources, 
 Active galactic Nuclei (AGN), Galaxy Clusters, Ultra High Energy Cosmic rays, Blazars and Star forming Galaxies, and (c) Resolved Point Sources (RPS),
 which constitute an  important part of photon background from  the direction of the Galactic Centre. The current sensitivity of the FERMI satellite 
 data is unfortunately hidden by the above background components, especially if uncertainties in the effective area of the detector are taken into account.
The situation will hopefully improve in the next few years, with the reduction of systematic errors. However, as the analysis of  ref.~\refcite{spanosgamma} 
demonstrates, the prospects for CMSSM LSP  indirect detection are not great. In particular,  in the low tan$\beta$ case, it will be 
 very difficult to detect a $\gamma$ Ð ray signal along
the co-annihilation strip but the focus point region has better prospects of detection due to the larger annihilations at that region.  On the other 
hand, better prospects seem to characterise the large tan$\beta$ case, due to  larger annihilation cross section in 
the co-annihilation, funnel and focus point regions.  In general, it will always be more difficult to pin down the CMSSM than other
supersymmetric models via searches for energetic photons from 
astrophysical sources. Hence collider searches are much superior in this respect for falsifying CMSSM, mSUGRA models.

Another set of indirect DM tests is that of neutrinos produced as a result of 
Capture and Annihilation of LSP in the Sun. The dominant process for the production of neutrinos is the annihilation of the LSP into SM particles, mainly tau pairs, the subsequent decay of which produces neutrinos and muons, and it is the detection of muons that eventually provides the main indirect DM test: $ \chi - \chi \rightarrow {\overline \tau} \, \tau $,  $ \quad \tau \rightarrow \mu \, \nu_\mu \, {\overline \nu_\tau} $.
Muon detection energy threshold is an important parameter for these tests. The fluxes of neutrinos (and hence muons) have been calculated 
again for CMSSM in ref.~\refcite{spanosnu}, at the same benchmark points as for the $\gamma$-ray spectra, with the conclusion that the detectability of CMSSM by the ICE CUBE/DEEP CORE detector is not straightforward, given that the signal above the background depends on the shape of the neutrino spectrum. The detailed mechanism for the production of detectable neutrino and thus muon flux from the Sun is the following: in the beginning we have  the Gravitational Capture of LSP from a  galaxy by the Sun,  then the LSP scatters off a nucleus in the Sun, 
loses energy and is  captured (as it cannot escape Sun's gravitational potential). Further scatterings in the Sun during the LSP's fall towards the solar centre  take place,  resulting in thermalization  (equilibrium situation) at the Solar Centre.  The increase of thermalized LSP populations  implies an   
increase in  the LSP annihilation rates. There is significant dependence of the calculated LSP annihilation rates on the solar model used as well as the particle physics model (spin dependent and spin independent cross sections). 
Indirect Searches for LSP DM via annihilations yielding high 
energy neutrinos (and hence muons) is \emph{not the most promising }
route for discovering SUSY, at least within CMSSM. But, as the detailed analysis of  refs.~\refcite{spanosnu,spanosnu2} has indicated,  there are 
models beyond the CMSSM, such as the Non Universal Higgs Mass variants of CMSSM, with sensitivity close to 
that of ICE CUBE/DEEP CORE. In such models, ICE CUBE/DEEP CORE friendly fluxes there are in 
regions where the LSP has significant Higgsino component, which implies larger LSP masses as compared to the corresponding CMSSM case
along the focus point regions.  
\begin{figure}
\begin{center}
\psfig{file=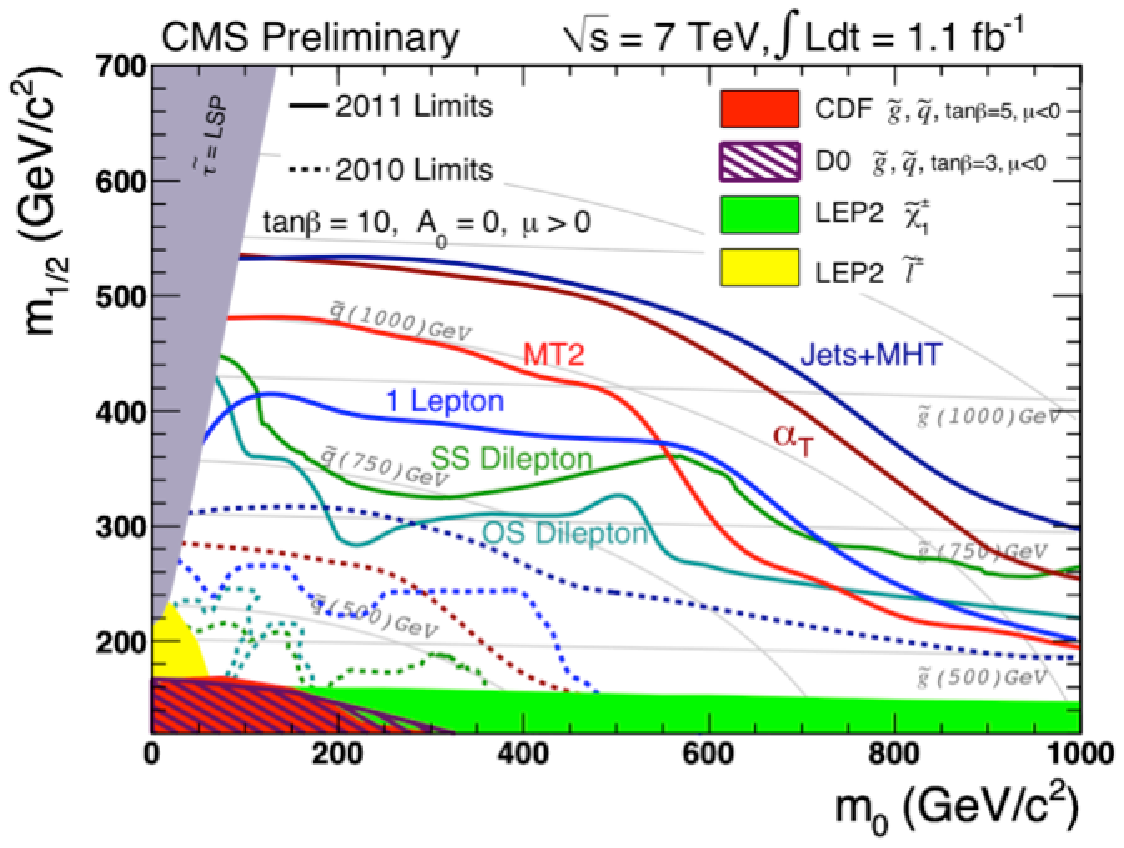,width=0.48\textwidth} \hfill
\psfig{file=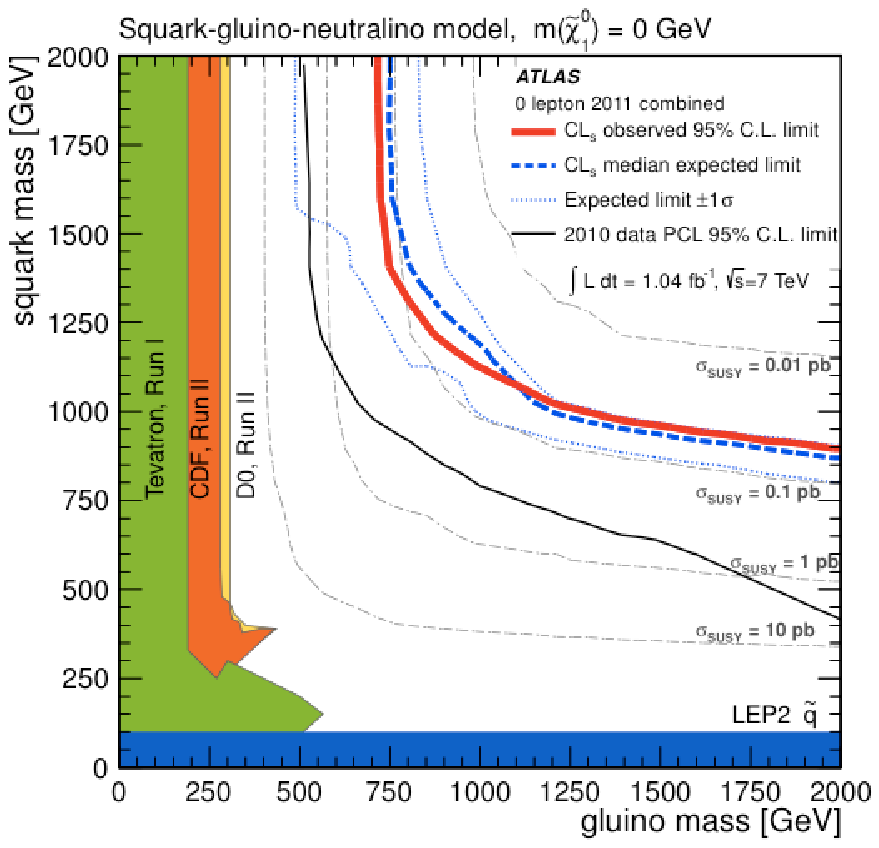,width=0.48\textwidth}
\end{center}
\caption{Current SUSY exclusion limits from the latest LHC results. Left panel : CMS Experiment, Right panel: ATLAS Experiment (ATLAS Collaboration, arXive:1109.6572)}
\label{fig:excl}
\end{figure} 

Unfortunately, at present there seem to be 
stringent exclusion bounds from the LHC  experiments (cf. fig. \ref{fig:excl}) for the CMSSM , mSUGRA models, excluding low SUSY partner masses of the type that constitute interesting regions of the CMSSM and its variants in the above indirect DM searches.  
Nevertheless, since the current exclusion limits from LHC (cf. fig. \ref{fig:excl}) pertain to missing transfer energy in interactions involving energetic jets and may be leptons, there are still regions of the parameter space that allow for minimal SUSY extensions of the SM: electroweak production, e.g. gaugino-gaugino production, compressed spectra (low sparticle mass differences, which imply low-momentum jets ) and 
third-generation sparticle production. Further exclusion will require improvement of systematic uncertainties, higher energies and relevant optimisation of analyses.   However, if the physical SUSY is realised through other models, then the conclusions may be completely different. Below we shall discuss two such departures  from mSUGRA.  The first concerns SUSY models  with broken R-parity, in which a long lived Gravitino (${\tilde G}$), with life time longer than the age of our Universe , plays the r\^ole of LSP. In some interesting variants of this class of models~\cite{valle}, with bilinear R parity Violation (RPV), neutrino masses are generated in an intrinsically supersymmetric way. The most promising (indirect detection) signal of the gravitino DM are monochromatic gamma-rays as a result of the gravitino decay modes ${\tilde G} \rightarrow \nu \, \gamma$, where $\nu$ indicates neutrinos. The model can be constrained  LHC data, neutrino oscillations, the WMAP astrophysical constraints on the relevant relic abundance 
$\Omega_\chi h^2$, $\gamma$-ray 
line searches (via Fermi, EGRET satellites). The allowed gravitino masses are below 1 GeV, with the corresponding life times longer than about 
$10^{28}$ sec. 

The second simplest class of models beyond the CMSSM, mSUGRA is provided by their coupling to cosmic time dependent scalar fields (dilatons) $\phi(t)$. The motivation for the use of such extensions  is that they entail relaxing to zero Dark Energy asymptotically  in cosmic time~\cite{lmn}, compatible with the current astrophysical data. Such couplings affect DM thermal species abundances,  as they modify the Boltzmann Equation by appropriate source terms dependent on the dilaton cosmic rates, $d\phi/dt$. The presence of the source and the associated corrections to $\Omega_\chi h^2$ may result in an O(10) dilution of the thermal relic abundance of the neutralino LSP in the CMSSM, while the baryon density remains unchanged. This results in more room for supersymmetry being available in the $(m_0\, m_{1/2}$) parameter space of the model~\cite{lmn},  compatible with the WMAP data, 
and thus heavier partners. The latter feature leads to new LHC signatures, for instance 
h($\rightarrow$ bb) + jets + MET,  Z($\rightarrow \ell \ell $) + jets + MET and  2$\tau$ + jets + MET, (MET= missing transverse energy)
are favoured in new regions. Such regions may be probed in the short future by the LHC detector. The fact that in such extensions of the CMSSM heavier partners are allowed, also implies larger annihilation cross section, and thus the above mentioned indirect DM signals via gamma rays and neutrinos from LSP annihilation have better prospects of detection, in comparison with the CMSSM case. However, the coupling of the dilaton to the relevant gravitino terms in the (conformal) supergravity Lagrangian will affect the gravitino decays rates, for instance ${\tilde G} \rightarrow \chi + Z^0$, where $\chi$ is the neutralino of the CMSSM, thus affecting the DM relic density and therefore implying stronger Big Bang Nucleosynthesis (BBN) constraints. It is therefore important that detailed cosmological studies of such dilaton extended mSUGRA/CMSSM models are performed~\cite{spnem}. 

A more complicated version beyond the mSUGRA comes from string theory on compactified manifolds. Such compactifications entail scalar fields with only gravitational couplings to ordinary matter (moduli), of which the above-mentioned dilaton is only one species. The stabilization of moduli is still an open issue in string theory, nevertheless there are consistent compactification schemes where such stabilization has been demonstrated consistently. One such example is the so-called G$_2$-MSSM, a stringy  extension of the minimal supersymmetric standard model in G$_2$ string manifolds~\cite{kane}, with hidden-sector induced moduli stabilization at TeV scale. The G$_2$-MSSM framework (which may or may not be characterised by an exact R-parity) gives rise to mostly Wino LSP (as opposed to mostly Bino LSP of the mSUGRA/CMSSM, where the  current LHC experimental constraints are placed (cf. fig.~\ref{fig:excl})). 
The predicted (physical) partner masses in this class of models are compatible with the corresponding current LHC exclusion bounds. Consistent cosmology also requires the moduli to dominate the energy density of the Universe before BBN, with a gravitino mass in the range 20-30 TeV. It is argued that a Wino-like LSP with large annihilation cross sections (of order $\langle \sigma v \rangle \sim 3 \times 10^{-24}$ cm$^3\, {\rm s}^{-1}$) and mass in the range 140-200~GeV arises from moduli decays prior to BBN and constitutes most of the (non-thermal) Dark Matter, consistently with current cosmological observations as well as direct dark matter searches. In particular, the model predicts~\cite{kane} spin independent cross sections which are  (at most of order 10$^{-45}$~cm$^2$) beyond the reach of the current XENON 100 Experiment, but  falsifiable in the next upgrade. 
\section{PAMELA/FERMI $e^+$ excess  and WIMPs \label{sec:pam}} 

Before proceeding to other DM candidates and their indirect searches, we should mention that the observed asymmetries between matter and antimatter by PAMELA, which have been confirmed by FERMI, in particular the observed positron excess in the Cosmic Ray (CR) spectra, but the absence of antiproton ($\overline{p}$) excess, can be accommodated~\cite{neutrpamela} within existing models of SUSY neutralino DM, although their most likely explanation may be astrophysical (pulsar emission~\cite{fermi}). Neutralino DM ineractions 
with SM particles produce charginos (next to lightest) , whose  
subsequent decay can produce  peak 
in the spectrum of cosmic leptons, yielding a 
signal analogous to that seen by PAMELA , ATIC and FERMI
(peak in the CR positron spectrum). 
The example studied in ref.~\refcite{neutrpamela} considered masses of neutralino   of order   110 GeV and chargino of order  250 GeV .
Such values are excluded by the current LHC data, and the question arises whether such scenarios survive the 
full LHC exclusion data, after four years of running.

However, in general, the PAMELA data may be compatible with generic WIMP DM. In particular, it has been argued in ref.~\refcite{strumia} that heavy 
($m_\chi \gg 1 {\rm GeV}$) DM annihilation can produce a jet structure which may result in antideuteron ($\overline{d}$) excess that can explain the lack of antriproton peak in the CR spectra, as observed by PAMELA and FERMI. The result seems pretty robust in the sense that astrophysical uncertainties do not affect significantly the ratio of concentrations $\overline{p}/ \overline{d}$ . The antideuteron signal is significantly enhanced for DM masses above 1 TeV. Moreover, in models where the Cosmology is modified, e.g.  by considering low reheating temperatures of the Universe after inflation in certain quintessence models~\cite{palis}, the relic density of DM WIMPs 
is found significantly enhanced compared to standard cosmology. In such models the calculated induced fluxes of $e^-, e^+$ in CR, produced by LSP annihilation into $e^- e^+, \, \mu^- \mu^+, \,  \tau^+ \tau^-$, indicate agreement with the results of PAMELA and FERMI as far as positron peak is concerned. Finally, we mention that the stringy moduli models of ref.~\refcite{kane}, characterised by Wino LSP,  are also compatible with the PAMELA data. 

Thus, although pulsar emission seems adequate to explain the current astrophysical data on observed ring of antimatter around the Earth, nevertheless 
several DM model explanations are also at play. 
Further astrophysical searches are therefore essential in order to settle this. In particular, if the pulsar explanation is the natural mechanism, then as we have heard in this meeting~\cite{fermi}, proton asymmetries in the CR spectra should be observed. This will hopefully be settled in the near future.

\section{ Sterile Neutrinos  as DM \label{sec:sterile}}  The SM CP violation cannot explain the observed matter-antimatter (baryon-antibaryon) asymmetry in the Universe. Several ideas beyond the SM (such as GUT models, Supersymmetry, Extra Dimensions \emph{etc}.) have been proposed in an attempt to resolve this issue. Right-handed supermassive neutrinos may provide extensions of SM with 
   extra CP Violation that can explain the origin of the observed matter-antimatter asymmetry in the Universe. 
   Such a scenario has been proposed in ref.~\refcite{shapo} as a non-supersymmetric minimal extension of the SM, called
 $\nu$MSM. The model may have several species N of right handed singlet Majorana neutrinos. 
The Model with one extra singlet fermion is excluded by the data, while models with 
2 or 3 singlet fermions work well in reproducing the Baryon Asymmetry and are consistent with experimental data on neutrino oscillations .
 In particular,  the Model with N=3 works fine, and in fact it allows one of the Majorana fermions  to almost decouple from the rest of the SM fields, thus providing a
candidate for light (kEV region of mass) sterile neutrino Dark Matter.  The other two right-handed neutrinos are degenerate in mass, and in fact much heavier than the third~\cite{shapo}, specifically one has:
Mass N$_2$ (N$_3$) / (Mass N$_1$)  = O(10$^5$ ), with the masses of N$_{2,3} < M_W = O(100) {\rm GeV}$ , with $M_W$ the electroweak symmetry breaking mass scale.
\begin{figure}
\begin{center}
\psfig{file=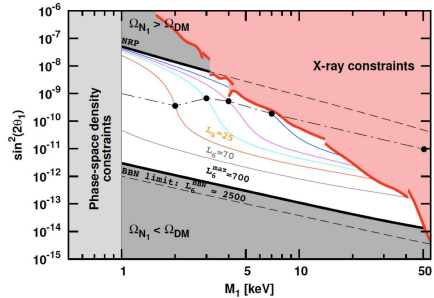,width=0.5\textwidth}
\end{center}
\caption{Astrophysical Constraints for the lightest sterile neutrino mass and mixing angle $\theta_1$ (with SM particles) in the $\nu$MSM model (from ref.~\refcite{shapo}). }
\label{fig:numsm}
\end{figure} 
The light neutrino masses are determined by the sea saw mechanism in the model. The lightest singlet neutrino is not stable, but its life time can be longer than the life time of the Universe, since its 
coupling  with the SM matter can be extremely weak . Under such conditions, its contributions to the mass of light neutrinos are well within experimental errors, and hence there is a consistent oscillation phenomenology of $\nu$MSM with light sterile neutrino  DM. Taking into account the interactions of the light neutrino with its heavier sterile partners in the $\nu$MSM, one may derive detailed constraints on the mass and couplings of the light sterile neutrino DM~\cite{shapo}. The reader should bear in mind that the decaying light sterile neutrino will produce narrow spectral lines in the spectra of DM dominated astrophysical objects, such as halos of galaxies etc. This will constitute a means of its detection. The $\nu$MSM model is found consistent with constraints from BBN , structure formation data in the universe and other astrophysical constraints. The allowed mass ranges and mixing angles $\theta_1$ are depicted in fig. \ref{fig:numsm}~\cite{shapo}.

\section{ Strongly Interacting Massive Particles (SIMPs) as DM candidates \label{sec:simp}} 

 Strongly interacting Massive particle (SIMP) matter may be (part) of DM, although much more severely constrained. 
 A rather old idea~\cite{simps} for a DM candidate is that the latter consists of 
 Charged Massive Particles (CHAMP). If the whole of DM, as originally assumed~\cite{simps}, consists of such charged particles, 
     then cosmological compatibilities require them  to be heavy, 20 TeV $<$  $M_{\rm Ch}$  $<$ 1000 TeV .
      Indeed, if of charge + 1, they will result in  Superheavy remnants of H isotopes in the Universe. CHAMPs are 
assumed  particle-antiparticle symmetric, so charge -1  anti-CHAMP may bind with $^4$He nuclei and after BBN. 
    Mostly, however, they bind to protons to behave like superheavy stable neutrons. Such bound states bring severe constraints in their relic populations.
Less severe constraints are imposed if CHAMPS constitute only (a small) part of DM: if neutral DM decays
(at late eras)  to CHAMPs then the above-mentioned stringent bounds may be re-evaluated,  for instance it has been estimated~\cite{champpartdm} that 
consistently with all current astrophysics constraints, the  fraction of CHAMP in the galactic halo may be less than  $0.4 Ð 1.4 \times  10^{-2}$. 
Also, it has been argued recently~\cite{kolb} that Galactic magnetic fields parallel to the disc prevent  CHAMPS from  entering the disc 
(hence their non detection on Earth), if their charge $q_X$  and mass are in the range: $10^2 (q_X/e)^2 \le m_X/(TeV)  \le  10^8 (q_X/e)^2$. Such CHAMPS
exert important influence on the DM density profiles: 
they interact with ordinary matter via magnetic field mediation and hence 
affect the visible Universe in the sense that their density profiles depend on the Galaxy: moderate effects appear in large elliptical galaxies and the Milky Way, while  there is expulsion of CHAMPS with moderate charge  (Coulomb Interactions 
not important) from spherical Dwarf Galaxies in  agreement with observations . 
Moreover, their DM Annihilation patterns are different from those of Cold Dark Matter (CDM) model:
due to the attractive Coulomb potential between X$^+$ and X$^-$ there is an 
increased annihilation cross section (relative to CDM models) 
by a factor $c/v$ (Sommerfield-Sakharov effect) ; 
after CHAMP becomes non relativistic,  the annihilation rate
 falls off slower than in CDM, their kinetic energies scale as (1 + z) with redshift z, and their 
present annihilation rate depends on the fraction of X$^-$ bound to baryons. 
  \begin{figure}
\begin{center}
\psfig{file=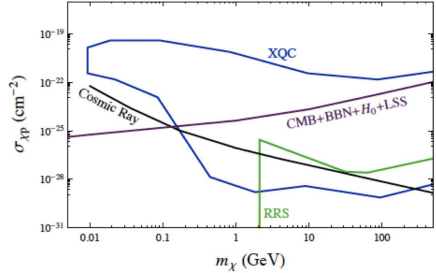,width=0.5\textwidth}
\end{center}
\caption{Current Astrophysical Constraints for neutral fermionic SIMP (from ref.~\cite{neutralsimp}). }
\label{fig:simp}
\end{figure}    

There are various experimental searches that impose stringent constraints on charged SIMPs:  as already said, the 
most important of them is associated with the formation of 
  \emph{Bound States SIMP-Nucleons}. Indeed,  if these are formed then the associated constraints exclude the models. 
  Hence one should 
avoid such bound states, e.g. by invoking repulsive forces between SIMPS and nucleons. 
To this end one may assume that SIMPS are 
fermionic and electrically neutral, since charged SIMPS of the same charge as nucleons (protons) could affect the Universe neutrality.  By assuming scalar field $\varphi$ mediators, as well as that the scalar force is less than that of two pions, so that bound states due to the scalar mediator do not form,  the pertinent part of the interacting Lagrangian is~\cite{neutralsimp}       
      $ L_{\rm int} = - g_X \varphi {\overline X} X - g_N \varphi {\overline N} N $, with $g_N g_X < 0, \quad m_X, m_\varphi > 0.$
 Important information is included in the SIMP(X) annihilation cross section $\sigma_{XX}$, which must be less than 
 $\sigma_{XX} \le 3 m_X {\rm GeV}^{-3}$ in order not to affect the shape of Galactic halo (the above upper bound is placed by an analysis of merging galaxies data, such as those of the Bullet Cluster~\cite{neutralsimp}). Such upper bounds allow for sufficiently strongly interacting particles. Another important constraint comes from  
 SIMP(X)-nucleon(N) cross section $\sigma_{XN}$ which must be less than  $\sigma_{XN} \le \frac{4 g_N^2}{g_X^2} \frac{m_X}{(1 GeV)} \times 10^{-27} cm^2$ so as not to affect the galactic halo shape. The resulting constraints from a plethora of astrophysical measurements, including X-ray quantum calorimetry (XQC) and Cosmic Rays  are depicted in fig. \ref{fig:simp}, indicating that light neutral fermionic SIMPs with masses less than 1 GeV are allowed by the current constraints.  An important collider signature of neutral fermionic SIMPS are DM di-jets produced in colliders, such as LHC. Indeed, as discussed in ref.~\refcite{neutralsimp}, the scattering length of a neutral fermionic SIMP $L_\chi = L_n \frac{\sigma_{\chi p}^{\rm inela}}{\sigma_{n p}^{\rm inela}  }$, where $n (p)$ indicates neutron (proton), can be smaller than the calorimeter size, so the SIMP can deposit energy in the form of DM jets. If the DM is neutral, then such jets would be trackless, and hence very different from the QCD one.        
          
\section{Other Interesting DM Possibilities \label{sec:otherDM}}

\emph{Axions }are also interesting candidates for DM, with a theoretical motivation, since their presence is associated with a resolution of the strong CP problem in QCD. In this talk, due to lack of time, I will not discuss them in detail. I will simply mention that the axion pseudoscalar field ${\tilde a}$ 
couples to the electromagnetic U(1) part of a GUT gauge group via terms of the form ${\tilde a} \vec{E} \cdot {\vec B}$ , where E and B are the electric and magnetic fields respectively. Such couplings imply that an axion field can be converted to a photon in the presence of an external magnetic field (Primakoff effect), which is the basis for their potential observation. The CAST experiment at CERN~\cite{cast} has placed the most stringent limits today to the QCD axions, given that no signal over background has been observed in the experiment. The preliminary limit on the axion couplings and masses are
$g_a < 2 - 2.5 \times 10^{-10} {\rm GeV}^{-1}$  for  the mass range 0.39 eV $ < m_a < $ 0.65 eV. The experiment, in addition to the QCD axions,  has also placed bounds on 
the couplings and compactification radius of Kaluza-Klein axions in extra dimensional theories.

In ref.~\refcite{khlopov}, \emph{Dark Atoms} from stable charged fundamental constituents 
                    of matter beyond  the Standard Model (new quarks and leptons) have been conjectured to exist. Severe constraints from anomalous isotopes
                    in the Universe  imply that only charge -2 object ($X^{--}$) is allowed, 
not charge +1 or  -1, hence the relevant models are necessarily  non supersymmetric. There are bound states of X$^{--}$ with primordial Helium 4 He$^{++}$ to form neutral  O-He atoms of warm DM. The non trivial (but unclear) nuclear physics of such bound states has been argued in ref.~\refcite{khlopov} to provide resolutions to various DM puzzles.  The O-He atoms may be responsible for the observed  constant and annual modulation of underground detectors  (like DAMA, CoGENT), while their decays to stable constituents may also explain the observed PAMELA and FERMI excess events. A clear signature of such models would be the appearance in the matter of DAMA detectors 
of anomalously heavy ($>$ 1 TeV)  Sodium Isotopes.

 Another interesting DM scenario has been presented in ref.~\refcite{hooper}. DM (which, in the model, is a right-handed sneutrino) communicates with SM particles via mediating light particles, e.g. Z$^\prime$ bosons of extra U(1)$^\prime$ groups that appear in extensions of the SM. Such Z$^\prime$ couple to the DM elastic cross sections, and can lead 
 (via DM annihilation) to excess
of $\gamma$-rays from, say,  the Galactic Centre, which are 
compatible with 
observations .

\vspace{-0.25 cm}
\section*{Acknowledgement} 

I thank J. Pinfold and the other organisers of the 13th ICATPP International Conference (Villa Olmo, Como (Italy), October 3-7 2011) for the invitation
to give this plenary talk and for organising such a stimulating event. This work was supported in part by the London Centre for
Terauniverse Studies (LCTS), using funding from the European Research
Council via the Advanced Investigator Grant 267352.


\vspace{-0.25 cm}
\begin{thebibliography}{99}

\bibitem{fermi} See plenary talks by F. Mocchiutti (PAMELA) and R. Rando (FERMI). 

\bibitem{fits} See, e.g.: A.~de la Cruz-Dombriz, V.~Gammaldi,
  %``Dark matter with photons,''
  [arXiv:1109.5027 [hep-ph]].

\bibitem{spanosgamma} J.~Ellis, K.~A.~Olive, V.~C.~Spanos,
  %``Galactic-Centre Gamma Rays in CMSSM Dark Matter Scenarios,''
  [arXiv:1106.0768 [hep-ph]].



\bibitem{spanosnu} J.~Ellis, K.~A.~Olive, C.~Savage, V.~C.~Spanos,
  %``Neutrino Fluxes from CMSSM LSP Annihilations in the Sun,''
  Phys.\ Rev.\  {\bf D81}, 085004 (2010).


 
\bibitem{spanosnu2}  J.~Ellis, K.~A.~Olive, C.~Savage, V.~C.~Spanos,
  %``Neutrino Fluxes from NUHM LSP Annihilations in the Sun,''
  Phys.\ Rev.\  {\bf D83}, 085023 (2011).

\bibitem{valle} D.~Restrepo, M.~Taoso, J.~W.~F.~Valle, O.~Zapata,
  %``Gravitino dark matter and neutrino masses with bilinear R-parity violation,''
  [arXiv:1109.0512 [hep-ph]].
  
\bibitem{lmn} A.~B.~Lahanas, N.~E.~Mavromatos, D.~V.~Nanopoulos,
  %``Smoothly evolving supercritical-string dark energy relaxes supersymmetric-dark-matter constraints,''
  Phys.\ Lett.\  {\bf B649}, 83-90 (2007); 
B.~Dutta {\it et al.},
 %``Supersymmetry Signals of Supercritical String Cosmology at the Large Hadron Collider,''
  Phys.\ Rev.\  {\bf D79}, 055002 (2009).

\bibitem{spnem} N.~E.~ Mavromatos and V.~C.~Spanos, to appear.


%\cite{Acharya:2011te}
\bibitem{kane}
  B.~S.~Acharya, G.~Kane, E.~Kuflik, R.~Lu,
  %``Theory and Phenomenology of $\mu$ in M theory,''
  JHEP {\bf 1105}, 033 (2011)  and references therein;
  %\cite{Acharya:2008zi}  
  B.~S.~Acharya, K.~Bobkov, G.~L.~Kane, J.~Shao, P.~Kumar,
  %``The G(2)-MSSM: An M Theory motivated model of Particle Physics,''
  Phys.\ Rev.\  {\bf D78}, 065038 (2008).

\bibitem{neutrpamela} A.~B.~Flanchik,
  %``Interaction of neutralino dark matter with cosmic rays and PAMELA/ATIC data,''
  [arXiv:1101.5920 [astro-ph.HE]].
  
  
\bibitem{strumia} M.~Kadastik, M.~Raidal, A.~Strumia,
  %``Enhanced anti-deuteron Dark Matter signal and the implications of PAMELA,''
  Phys.\ Lett.\  {\bf B683}, 248-254 (2010).
 

\bibitem{palis} C.~Pallis,
  %``Cold Dark Matter in non-Standard Cosmologies, PAMELA, ATIC and Fermi LAT,''
  Nucl.\ Phys.\  {\bf B831}, 217-247 (2010).
  



\bibitem{shapo} M.~Shaposhnikov,
  %``Baryon asymmetry of the universe and neutrinos,''
  Prog.\ Theor.\ Phys.\  {\bf 122}, 185 (2009) and references therein.
  %%CITATION = PTPKA,122,185;%%


  
\bibitem{simps} A.~De Rujula, S.~L.~Glashow, U.~Sarid,
  %``Charged Dark Matter,''
  Nucl.\ Phys.\  {\bf B333}, 173 (1990); 
  G.~D.~Starkman, A.~Gould, R.~Esmailzadeh, S.~Dimopoulos,
  %``Opening The Window On Strongly Interacting Dark Matter,''
  Phys.\ Rev.\  {\bf D41}, 3594 (1990).
  
\bibitem{champpartdm} F.~J.~Sanchez-Salcedo, E.~Martinez-Gomez, J.~Magana,
  %``On the fraction of dark matter in charged massive particles (CHAMPs),''
  JCAP {\bf 1002}, 031 (2010).
 
  
\bibitem{kolb} L.~Chuzhoy, E.~W.~Kolb,
  %``Reopening the window on charged dark matter,''
  JCAP {\bf 0907}, 014 (2009).


\bibitem{neutralsimp} Y.~Bai, A.~Rajaraman,
  %``Dark Matter Jets at the LHC,''
    [arXiv:1109.6009 [hep-ph]] and references therein.

\bibitem{cast} I.~G.~Irastorza,  {\it et al.},
  %``Latest results and prospects of the CERN Axion Solar Telescope,''
  J.\ Phys.\ Conf.\ Ser.\  {\bf 309}, 012001 (2011).

\bibitem{khlopov} M.~Y.~Khlopov,
  %``Dark Atoms of Dark Matter from New Stable Quarks and Leptons,''
  [arXiv:1012.5756 [astro-ph.CO]] and Poster here.  
 
\bibitem{hooper} M.~R.~Buckley, D.~Hooper, J.~L.~Rosner,
  %``A Leptophobic Z' And Dark Matter From Grand Unification,''
  Phys.\ Lett.\  {\bf B703}, 343-347 (2011); 
  M.~R.~Buckley, D.~Hooper, T.~M.~P.~Tait,
  %``Particle Physics Implications for CoGeNT, DAMA, and Fermi,''
  Phys.\ Lett.\  {\bf B702}, 216-219 (2011).
 
  
\end{thebibliography}
\end{document}